\documentclass[%
 reprint,
 amsmath,amssymb,
 aps,
]{revtex4-2}

\usepackage{graphicx}
\usepackage{dcolumn}
\usepackage{bm}
\usepackage{xcolor}

\begin{document}

\preprint{APS/123-QED}

\title{Optimizing Particle Transport for Enhanced Confinement in Quasi-Isodynamic Stellarators} 

\author{A.~Ba\~n\'on~Navarro}

\author{A.~Di Siena}

\author{F.~Jenko}
\affiliation{Max Planck Institute for Plasma Physics, Boltzmannstr.~2, 85748 Garching, Germany}
\author{A.~Merlo}
\author{E.~Laude}
\affiliation{Proxima Fusion, Flößergasse 2, 81369 Munich, Germany}

\date{\today}

\begin{abstract}

Despite significant advances in reducing turbulent heat losses, modern quasi-isodynamic (QI) stellarator continue to suffer from poor particle confinement, which fundamentally limits their overall performance. Using gyrokinetic simulations within the GENE–Tango framework, we identify suppressed inward thermodiffusion, caused by unfavorable magnetic geometry, as the primary cause. To overcome this limitation, we design a new configuration with a reduced mirror ratio, which enhances the contribution of passing electrons to the inward particle flux. This facilitates the formation of strongly peaked density profiles, suppresses turbulence, and leads to a substantial improvement in confinement. Our optimized configuration achieves nearly a twofold increase in energy confinement compared to \textit{Stellaris}, highlighting the crucial role of optimizing particle transport in next-generation stellarator designs.

\end{abstract}

\maketitle


Understanding turbulent transport in stellarator devices has become increasingly important, especially in light of the substantial progress made over the past decades in optimizing neoclassical transport. Wendelstein 7-X (W7-X) exemplifies this advancement, having achieved significant reductions in neoclassical transport through careful magnetic configuration optimization, with turbulent transport now limiting overall confinement~[\onlinecite{Wolf_2017, Klinger_2019, Beidler2021}]. This success has led to a paradigm shift in stellarator design, where turbulent transport, previously neglected in optimization metrics, is now being actively included. Consequently, various optimization strategies have emerged, including minimizing flux compression in regions of unfavorable magnetic curvature~[\onlinecite{Stroteich_Xanthopoulos_Plunk_Schneider_2022},\onlinecite{PRXEnergy.3.023010}], increasing the critical gradient for turbulence stabilization~[\onlinecite{PhysRevResearch.5.L032030}, \onlinecite{Roberg-Clark_Xanthopoulos_Plunk_2024}], reducing magnetic curvature itself~[\onlinecite{PhysRevLett.105.095004, 10.1063/1.3560591, Mynick_2014}], optimizing max‑J configurations that confine trapped particles to regions of favorable curvature~[\onlinecite{Velasco_2023, Sanchez_2023, García-Regaña_2025, PhysRevLett.133.185101}], and other approaches aimed at mitigating turbulent transport~[\onlinecite{Kim_Buller_2024, PhysRevE.110.035201, Guttenfelder_Mandell_2025}]. These turbulence optimization methods also require retaining essential features necessary for stellarator reactors, such as strong neoclassical and energetic particle confinement and reduced bootstrap current~[\onlinecite{Helander_2009, PhysRevLett.128.035001, doi:10.1073/pnas.2202084119, Sánchez_2023, Goodman_Camacho}], MHD stability~[\onlinecite{Helander_2014}, \onlinecite{Rodríguez_Mackenbach_2023}], favorable max-J properties~[\onlinecite{PhysRevLett.108.245002, Proll_2022, Rodríguez_Helander_Goodman_2024}], and manageable coil complexity~[\onlinecite{doi:10.1073/pnas.2202084119}, \onlinecite{ Kappel_2024}]. An example of this approach is the development of the quasi-isodynamic {\textit{SQuID}} stellarator configurations~[\onlinecite{PRXEnergy.3.023010}, \onlinecite{Goodman_Camacho}]. Their optimized magnetic geometry ensures excellent fast ion and neoclassical confinement, along with MHD stability, while simultaneously reducing turbulent transport.

These stellarator configurations—including the latest design for \textit{Stellaris}, a fusion power plant concept by Proxima Fusion, the Max Planck Institute for Plasma Physics, and other academic partners —exhibit reduced turbulent transport, particularly in terms of heat fluxes~[\onlinecite{LION2025114868}]. However, a detailed understanding of particle transport in these highly optimized geometries remains an open question. Accurately predicting particle transport is essential for evaluating plasma performance, as it is closely coupled to heat fluxes through the plasma density profile—for example, via density gradient stabilization of ion temperature gradient (ITG) turbulence. To address this, we present the first high-fidelity turbulence simulations of \textit{Stellaris} with evolved density profiles, utilizing gyrokinetic flux-matching simulations performed with the \texttt{GENE} code~[\onlinecite{Jenko_POP_2010}] coupled to the~\texttt{Tango} transport solver~[\onlinecite{Parker_2018}]. The GENE--Tango framework has been extensively validated in tokamak geometries across a variety of scenarios at ASDEX Upgrade~[\onlinecite{Di_Siena_2022}, \onlinecite{Di_Siena_2024}] and JET~[\onlinecite{Di_Siena_2025}], consistently demonstrating excellent agreement with experimental measurements. More recently, it has also been validated against four experimental scenarios on W7-X~[\onlinecite{don_fernando_2025}], establishing it as one of the most reliable tools currently available for predictive performance studies in stellarator configurations.

\begin{figure}[t]
\centering
\includegraphics[width=\linewidth]{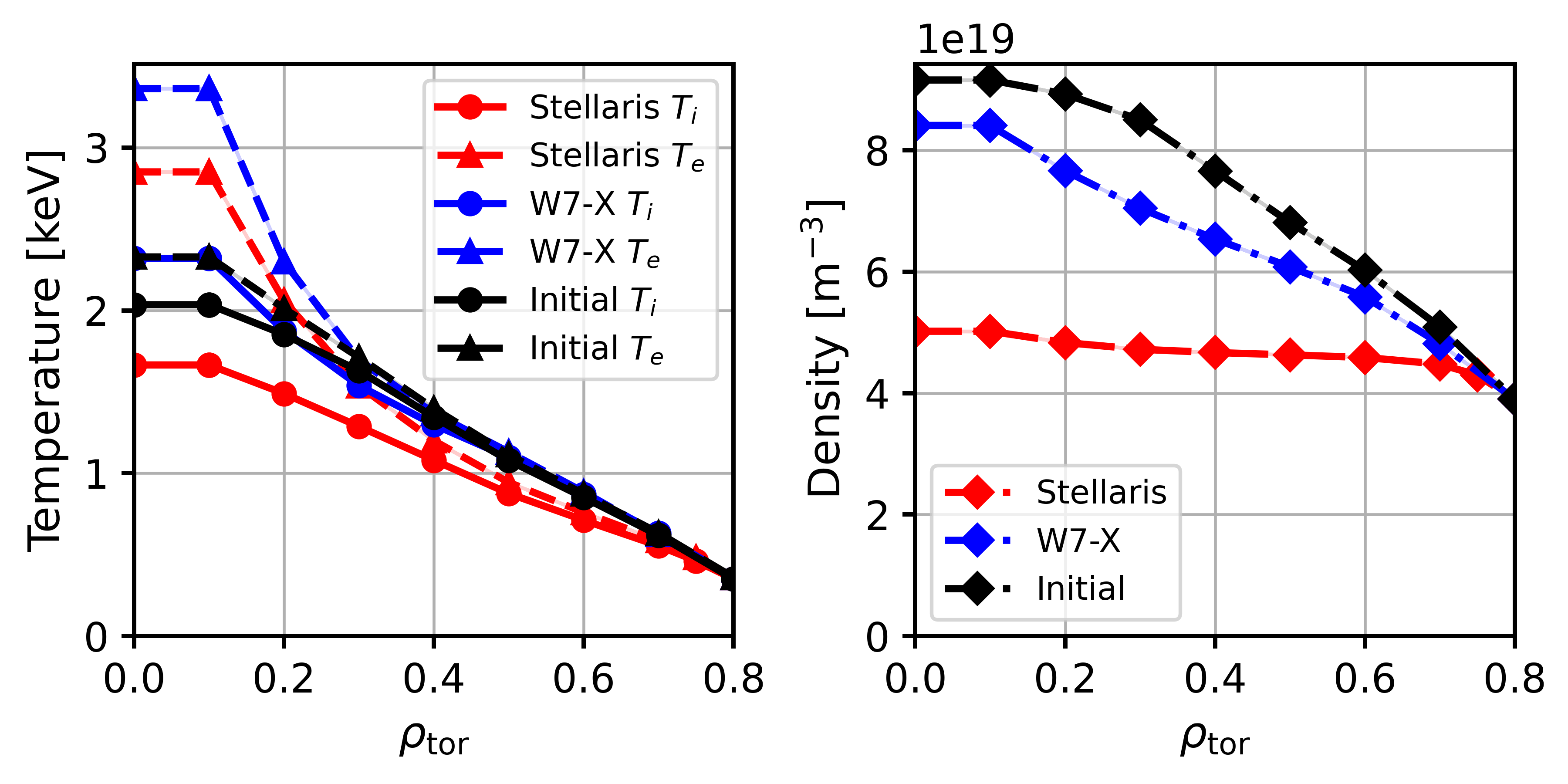}
\caption{ Left: Ion and electron temperature profiles ($T_i$, $T_e$); Right: Corresponding density profiles — all plotted as functions of normalized radius $\rho_{\mathrm{tor}}$. GENE flux-tube simulations are performed along the magnetic field line defined by $\alpha = q \theta^{\star} - \phi = 0$, with $\rho_{\rm tor}$ ranging from $0.1$ to $0.8$ in steps of $0.1$ for W7-X. For \textit{Stellaris}, an additional point at $\rho_{\rm tor} = 0.75$ is added to improve convergence near the boundary. Here, $q$ is the safety factor, $\theta^{\star}$ is the PEST poloidal angle, and $\phi$ is the toroidal angle. Simulations use a resolution of $(128, 64, 192, 32, 12)$ in $(k_x, k_y, z, v_{\parallel}, \mu)$, where $k_x$ and $k_y$ are the radial and binormal wavenumbers, $z$ is the coordinate along the field line, $v_{\parallel}$ is the parallel velocity, and $\mu$ is the magnetic moment. The radial box size is $l_x \approx 128 \, \rho_i$, and the minimum binormal wavenumber is $k_{y,\rm min} \, \rho_i = 0.05$. We use generalized twist-and-shift boundary conditions~[\onlinecite{Martin_2018}] in the parallel direction with $n_{\rm pol} \approx 1$–$1.5$, depending on radius and configuration. The parallel velocity box size is $l_{v_{\parallel}} = 3 \sqrt{2} v_i$, and for the magnetic moment, $l_{\mu} = 9 \, T_i / B_0$.
}
\label{fig:profile_comparision}
\end{figure}

To benchmark \textit{Stellaris}, we compare its performance to a reference W7-X plasma from the GENE--Tango validation study~[\onlinecite{don_fernando_2025}] of gas puffing fueling ECRH discharge (1MW) with a particle source of $4.5\times10^{20},\mathrm{s^{-1}}$ plus neutral beam injection (3MW). 
The boundary conditions for ion and electron temperatures and density are fixed at the normalized toroidal flux radius $\rho_{\mathrm{tor}} = 0.8$, where $\rho_{\mathrm{tor}} = \sqrt{\Phi/\Phi_{\mathrm{edge}}}$ is defined as the square root of the ratio of enclosed toroidal magnetic flux $\Phi$ to the total edge flux $\Phi_{\mathrm{edge}}$. From this location, plasma profiles are evolved inward to the magnetic axis until a steady state is reached, wherein turbulent transport matches the integrated external sources across all modeled channels and radii. For a fair comparison, we retain the plasma size and magnetic field strength from W7-X in both geometries, with a minor radius of $a = 0.5$ m and an on-axis magnetic field strength of $B_0 = 2.6$ T. The resulting steady-state temperature and density profiles are shown in the top row of Fig.~\ref{fig:profile_comparision}.

Surprisingly, we observe that despite \textit{Stellaris} being specifically optimized to reduce turbulent transport, it achieves significantly lower plasma pressure compared to W7-X. This reduction is primarily attributed to a substantial degradation in particle confinement.  As a result, while W7-X maintains a strong density gradient across the radial domain, \textit{Stellaris} exhibits a marked collapse in plasma density. This density collapse has a secondary impact on the temperature profiles. With the density gradient largely vanishing, \textit{Stellaris} loses the stabilizing effect that finite density gradients exert on ITG turbulence. Consequently, the temperature profiles in \textit{Stellaris} become broader and less peaked than those in W7-X.

These findings suggest that the current optimization pathway—focused primarily on reducing turbulent heat transport—overlooks a critical ingredient: the control of particle confinement and, in particular, the density profile. As highlighted in recent W7-X experimental campaigns~[\onlinecite{Xanthopoulos_PRL_2020, Laqua_2021, Ford_2024}], maintaining a high core density is essential for achieving improved overall performance. This consideration becomes even more important for future stellarator reactors, where alpha particle heating scales with the core density. Devices capable of sustaining higher densities at similar temperatures may enter a positive feedback loop, where increased alpha heating further boosts central plasma pressure, ultimately enabling higher fusion performance.


\textit{Can We Control Density Build-Up?} In this Letter, we present the first design strategy for stellarator configurations that simultaneously achieve exceptional particle and heat confinement. This is accomplished by combining established turbulence optimization methods with new insights into the physics of particle transport. In particular, we enhance the contribution of inward particle flux, enabling the formation of more peaked density profiles.

The underlying optimization strategy can be understood by analyzing the key dependencies of particle flux, assuming turbulence locality. This assumption is well supported by current experiments and is expected to hold even more strongly in reactor-relevant regimes. Under this approximation, the electron particle flux $\Gamma_e$ can be decomposed into diffusive, thermodiffusive and convective components as follows:
\begin{equation}
\Gamma_e = 
D_e \frac{a}{L_{n,e}} + D_{T,e} \frac{a}{L_{T,e}}  + a V_{p,e},
\end{equation}
with, \( D_{e} \), \( D_{T,e} \) and \( V_{p,e} \) respectively the diffusion, thermodiffusion and convective transport coefficients, as defined in Ref.~[\onlinecite{10.1063/1.3155498}], and $a/L_{n_e}$ and $a/L_{T_e}$ are the normalized electron density and temperature gradients, respectively, defined as $a/L_X \equiv -a\, d\ln X/dr$, with $X$ representing either $n_e$ or $T_e$. For completeness, and given the central role of the thermodiffusion coefficient, we provide its definition below:
\begin{align}
D_{T,e} &= \frac{k_y c_s^2}{\Omega_{ci}} 
\left\langle \int d^3v\, F_M 
\frac{ \left(\gamma_k + \hat{\nu}_k \right)\left[ \frac{v^2}{T_e} - \frac{3}{2} \right]}
{ \left( {\omega}_k + \hat{\omega}_d \right)^2 
+ \left( \gamma_k + \hat{\nu}_k \right)^2 } \right. \notag \\
&\qquad\left. \times\, J_0^2(k_\perp \rho_s)\, |\hat{\phi}_k|^2
\right\rangle.
\end{align}
Here, \( \gamma_k \) and \( \omega_k \) are the normalized growth rate and real frequency of the unstable linear mode, respectively. The normalized electrostatic potential is \( \hat{\phi}_k = e \phi_k / T_e \), and the normalized magnetic drift frequency is \( \hat{\omega}_d = (v_\parallel^2 + v_\perp^2 / 2) / v_{\mathrm{th}e}^2 \, \hat{\omega}_{Dk} \), where \( v_{\mathrm{th}e} = \sqrt{2 T_e / m_e} \) is the electron thermal speed. The magnetic drift frequency is defined as \( \omega_{Dk} = k_y \mathcal{K}_y c_s / a \), where \( k_y \) is the binormal wavenumber, \( \mathcal{K}_y \) is the binormal curvature, and \( c_s = \sqrt{T_e / m_i} \) is the ion sound speed, with \( T_e \) the electron temperature and \( m_i \) the ion mass. The normalized collision frequency is \( \hat{\nu}_k = \nu_{ei} v_{\mathrm{th}e}^3 / (v^3 \omega_{Dk}) \), with \( \nu_{ei} \) the electron-ion collision frequency. The total particle speed is defined as \( v^2 = v_\parallel^2 + \mu B_0 \). \( F_M \) denotes the Maxwellian distribution function, \( J_0 \) is the zeroth-order Bessel function of the first kind, and \( k_\perp \rho_s \) is the normalized perpendicular wavenumber, with \( \rho_s = c_s / \Omega_{ci} \) the ion sound gyroradius and \( \Omega_{ci} = e B_0 / m_i \) the ion gyrofrequency. This expression neglects parallel motion and assumes a simple Krook collision operator, whereas the GENE simulations retain the full dynamics and employ a Landau–Boltzmann collision operator.

The diffusion term is always positive, representing outward transport. The thermodiffusive term is proportional to \( (v^2/T_e - 3/2) \), leading to a larger outward flux of particles with large velocities and a compensating inward flux of slower particles. The resulting net thermodiffusive flux is determined by the balance of these contributions. Finally, the convective term is a purely geometric effect, arising from magnetic field curvature and inhomogeneities, and is often referred to as the curvature pinch. It can be directed either inward or outward depending on the mode and geometry.

Our strategy aims to enhance inward thermodiffusive transport by increasing the contribution of slow electrons—those with normalized velocities \( v^2/T_e < 3/2 \)—which preferentially drive inward flux. This is achieved by combining two elements: increasing collisionality to suppress the trapped-particle population via de-trapping, and tailoring the magnetic geometry to reduce the mirror ratio, thereby increasing the passing-particle fraction. Together, these effects amplify the role of passing electrons in driving inward transport. This mechanism is clearly reflected in our numerical results.

\begin{figure}[!htb]
    \centering
    \begin{minipage}[b]{0.48\columnwidth}
        \centering
        \includegraphics[width=\linewidth]{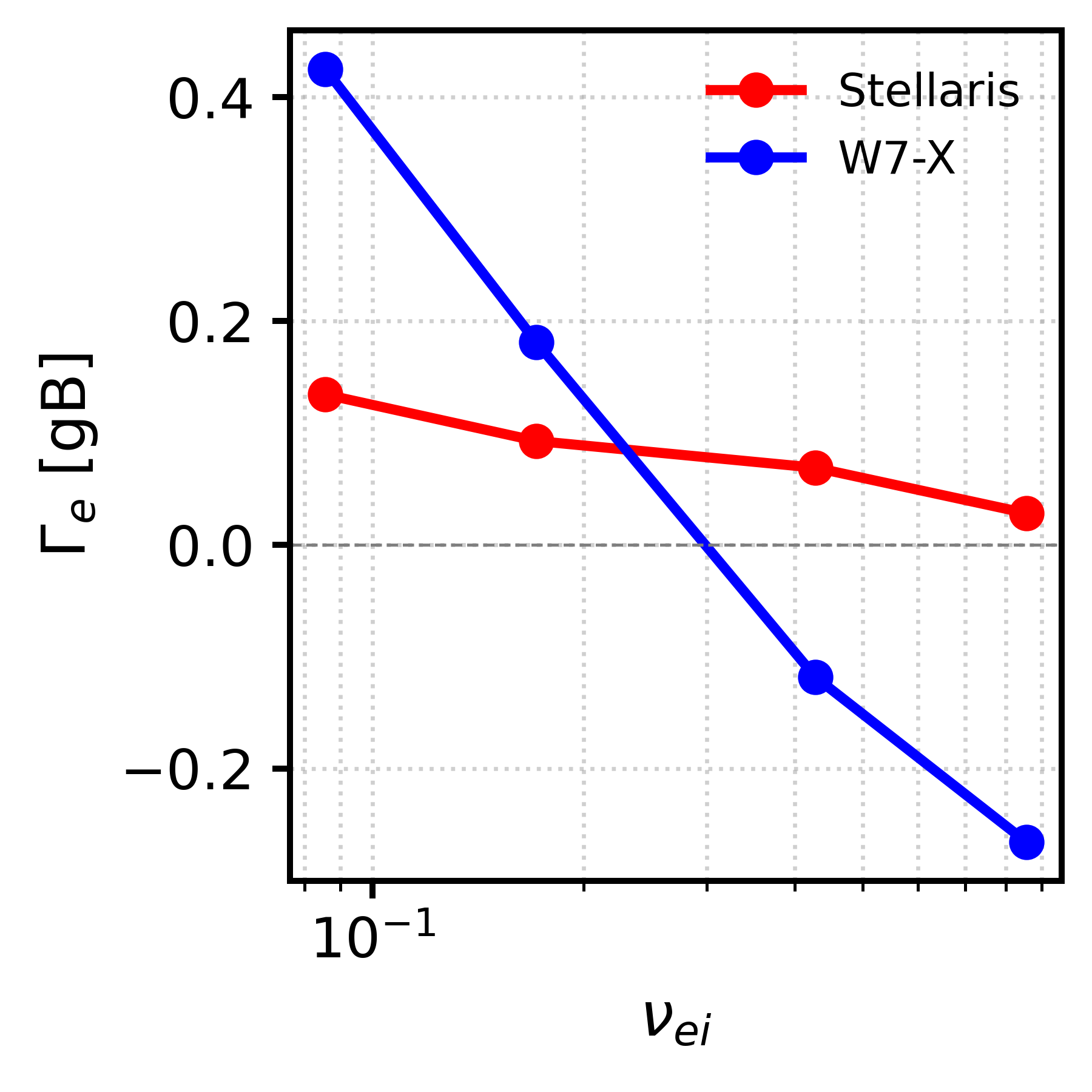}
    \end{minipage}
    \hfill
    \begin{minipage}[b]{0.48\columnwidth}
        \centering
        \includegraphics[width=\linewidth]{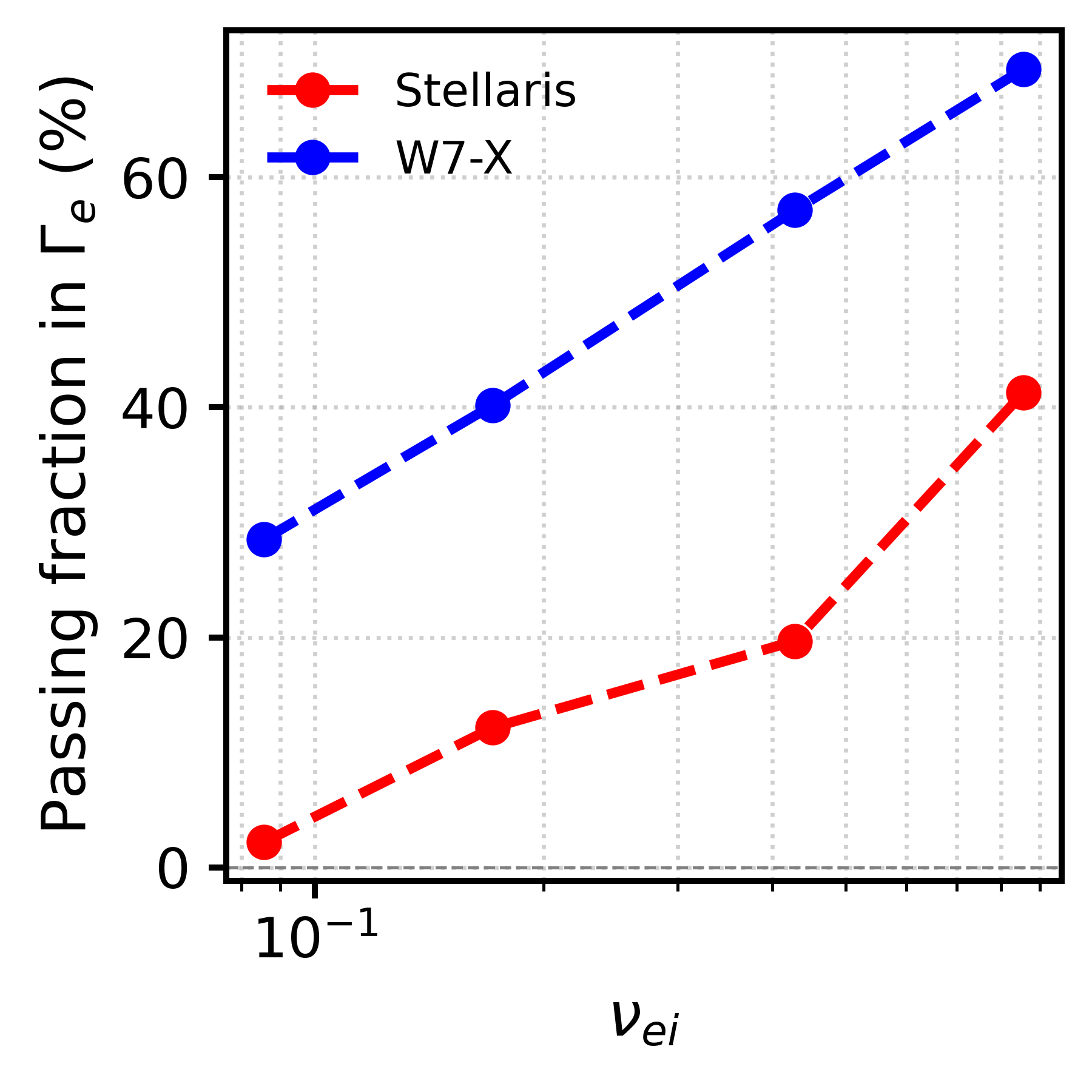}
    \end{minipage}
    \vspace{0.5em}

    \begin{minipage}[b]{0.98\columnwidth}
        \centering
        \includegraphics[width=\linewidth]{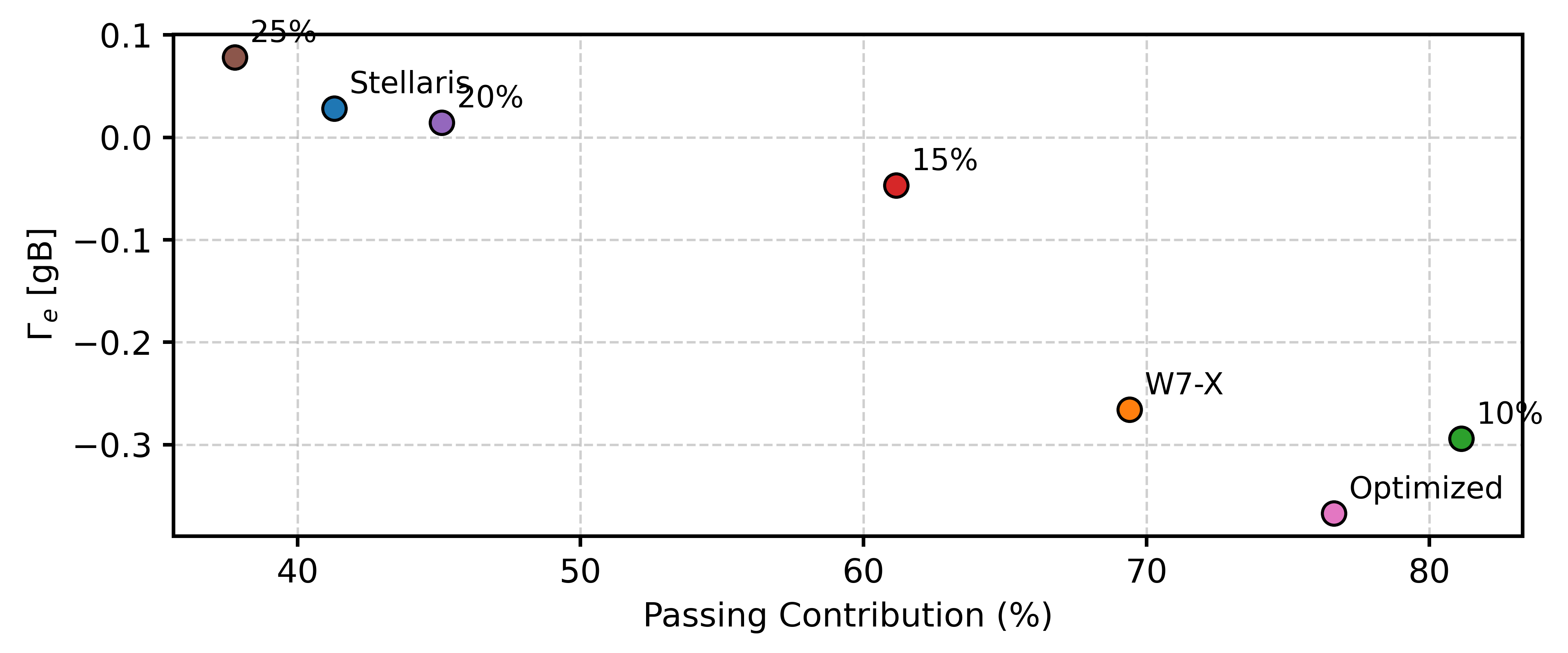}
    \end{minipage}
    \caption{Top left: The total electron particle flux is expressed in gyro-Bohm units, \([ \text{gB} ] = n_e c_s \rho_s^2/a^2 \),   as a function of collisionality ($\nu_{e,i}$) for values representative of the profiles shown in Fig.~1, comparing W7-X (blue) and \textit{Stellaris} (red). 
    Top right: Decomposition of the particle flux into passing and trapped electron contributions for the same configurations and collisionalities. Bottom: Inward particle flux versus mirror ratio across a series of configurations with varying on-axis mirror ratios (expressed in \%). The \textit{Optimized} configuration is designed to follow the radial mirror ratio profile of W7-X. A strong correlation is observed: reducing the mirror ratio increases the passing-particle contribution and enhances the inward pinch. GENE flux-tube simulations were performed at $\rho_{\rm tor} = 0.8$ using the same resolution as in the GENE-Tango simulations, but with profile parameters such as $a/L_{T,e} = a/L_{T,i} = 2.2$ and $a/L_{n} = 0.5$, and $T_e/T_i = 1$.
    }
    \label{fig:triple_panel}
\end{figure}

Figure~\ref{fig:triple_panel}, top left, shows how increasing collisionality influences particle transport in both W7-X and the \textit{Stellaris} configuration. A clear trend emerges: the total particle flux decreases with collisionality. However, while W7-X exhibits a robust inward pinch, \textit{Stellaris} saturates near zero flux and does not enter the pinch regime. To understand this contrast, we decompose the total flux into contributions from passing and trapped particles. As shown in Figure~\ref{fig:triple_panel}, top right,, the inward flux in W7-X is predominantly carried by passing electrons, whose contribution is significantly larger than in \textit{Stellaris}. This difference is directly linked to the magnetic geometry: W7-X's lower mirror ratio (approximately 3.5\% on-axis) yields a greater passing-particle fraction, while \textit{Stellaris}’s higher mirror ratio (approximately 25\% on-axis) results in a larger trapped population that contributes less effectively to inward transport. Thus, reducing the mirror ratio and increasing collisionality act in synergy to enhance inward thermodiffusion via passing electrons.

This trend is further illustrated in Figure~\ref{fig:triple_panel}, bottom, which shows the particle flux for a set of magnetic configurations generated by systematically modifying the magnetic geometry to reduce the mirror ratio—from approximately 25\% to 10\%—while keeping other parameters fixed. For each configuration, the figure also displays the fraction of the total particle flux carried by passing electrons. The results clearly demonstrate that lowering the mirror ratio significantly strengthens the inward particle pinch, with a growing share of the flux attributed to passing electrons. This behavior is robust: it holds across different radial positions and persists over the full range of collisionalities considered.


\textit{New density-optimized configuration.} Building on these insights, we propose a new density-optimized magnetic configuration designed to enhance inward thermodiffusion. This configuration has a significant lower overall mirror ratio compared to \textit{Stellaris}. Consequently, it increases the population of slow passing electrons and maximizes the inward particle flux—particularly at the higher collisionalities relevant for future fusion devices.

\begin{figure}[t]
\centering
\vspace{-2.0cm}
\includegraphics[width=\linewidth]{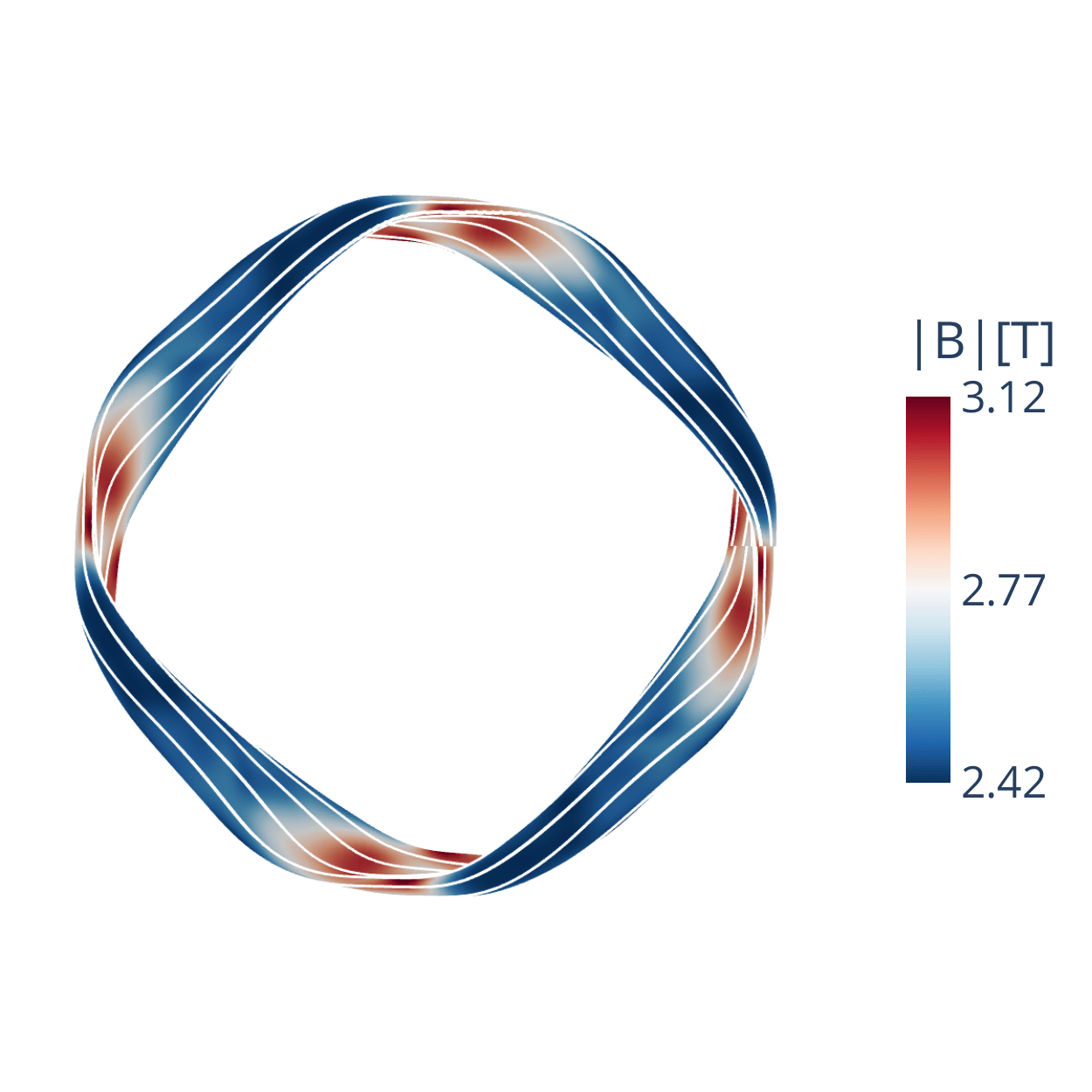}
\vspace{-2.0cm}
\caption{
    The magnetic field strength at the plasma boundary of the {\it Optimized} configuration. White lines represent magnetic field lines.
}
\label{fig:plasma_boundary}
\end{figure}

This corresponds to the case shown in Figure~\ref{fig:triple_panel}c (\textit{Optimized}), which exhibits the strongest negative particle flux. The new design aims to improve particle confinement and, consequently, suppress heat transport by producing strongly peaked density profiles. These profiles stabilize ITG turbulence through density gradient stabilization, leading to significantly enhanced overall performance. The {\it Optimized} configuration is shown in Figure~\ref{fig:plasma_boundary}. 
The optimization targets a configuration with aspect ratio $A^{\ast}=11.5$,
on-axis magnetic mirror ratio $\Delta^{\ast}=5\%$,
a vacuum magnetic well $W_{vacuum}^{\ast}=1\%$.
The rotational transform is constrained within $\iota \in [0.8, 1.0]$,
and the flux surface compression in regions of unfavorable curvature is kept below a reference value.
Moreover, we minimize the deviation from a QI field at mid-radius and at the plasma boundary following~[\onlinecite{PRXEnergy.3.023010}].
We solve the constrained optimization problem using an Augmented Lagrangian method,
employing CMA-ES as the inner oracle~[\onlinecite{hansen2001completely},\onlinecite{cadena2025constellaration}].

We assess neoclassical transport in the $1/\nu$ regime using the \texttt{NEO} code~[\onlinecite{nemov1999evaluation}] to compute the effective ripple,
$\epsilon_{\text{eff}}$,
which ranges from $0.1\%$ at the magnetic axis to $1.2\%$ at the plasma boundary.
The configuration is stable according to the Mercier criterion and it possess negative infinite-$n$ ballooning growth rates up to a volume-averaged beta of $\langle \beta \rangle_V \simeq 2.0\%$.
Alpha-particle confinement is evaluated using the \texttt{SIMPLE} code~[\onlinecite{albert2020accelerated}].
We simulate the trajectories of 5000 fusion-born $\alpha$-particles with 3.5~MeV kinetic energy,
uniformly distributed in pitch angle and isotropically launched at mid-radius.
Each trajectory is followed for 0.1~s.
As is standard practice for this assessment,
the configuration is scaled to the ARIES-CS design point~[\onlinecite{najmabadi2008aries}].
Confinement improves with increasing plasma beta:
the loss fraction decreases from $20\%$ at $\langle \beta \rangle_V = 2\%$ to $15\%$ at $\langle \beta \rangle_V = 4\%$.

\begin{figure*}[!htb]
\centering
\includegraphics[width=\textwidth]{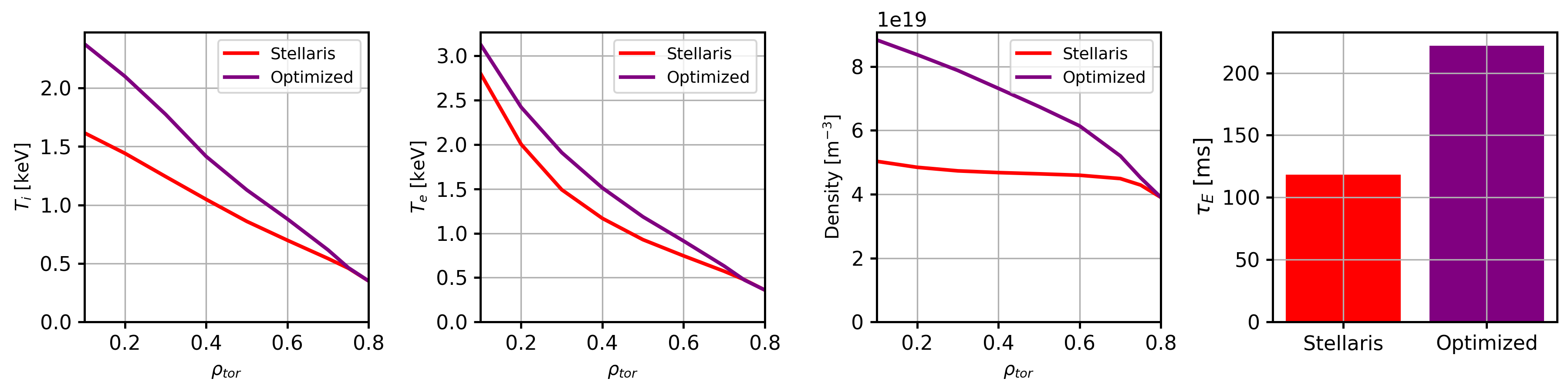}
\caption{Comparison of plasma profiles for the density-optimized configuration  (purple) and \textit{Stellaris} (red). Shown are the ion temperature \(T_i\), electron temperature \(T_e\), and density \(n\) profiles. The fourth figure presents the calculated energy confinement times in milliseconds [ms], demonstrating the performance improvement of the optimized configuration.}
\label{fig:profile_comparision_all}
\end{figure*}

The performance of our density-optimized configuration, is evaluated using GENE–Tango with the same setup employed for \textit{Stellaris}. This ensures a fair comparison by matching the minor radius and magnetic field on axis. The results are presented in Figure~\ref{fig:profile_comparision_all}, demonstrating a substantial improvement across all modeled plasma profiles. In particular,  the optimized configuration achieves nearly a twofold increase in calculated energy confinement time compared to \textit{Stellaris}.
%


\textit{Summary and discussion.} In this Letter, we have identified particle confinement as a key performance bottleneck in modern turbulence-optimized stellarators—a factor largely overlooked in previous optimization efforts. Using advanced gyrokinetic simulations within the GENE–Tango framework, we demonstrate that this limitation can be overcome by maximizing the inward particle flux driven by thermodiffusion. We design a new magnetic configuration with a significantly reduced mirror ratio compared to existing turbulence optimized designs, thereby increasing the fraction of passing electrons responsible for the inward particle pinch. This enables the formation of strongly peaked density profiles.

The resulting density gradient stabilizes ITG turbulence, leading to substantial improvements in both particle and heat confinement. Our density-optimized configuration achieves nearly a twofold increase in energy confinement relative to \textit{Stellaris}, under identical plasma parameters and heating conditions. These results highlight the critical role of particle transport optimization as a necessary complement to traditional approaches focused solely on turbulent heat flux reduction.

Future stellarator designs should incorporate particle transport optimization as a central objective—alongside turbulence reduction—to achieve the high core densities and robust confinement required for reactor-relevant performance. 
The optimization strategy proposed in this Letter offers a promising route toward this goal, enabling simultaneous control of heat and particle transport to surpass current performance limits. Our results strongly motivate continued theoretical and experimental investigations into particle flux control and the development of magnetic geometries that exploit key physical mechanisms for improved confinement and enhanced fusion performance.

Beyond the strategy presented in this Letter, additional pathways might be explored to further improve particle transport. For instance, the convective particle flux—strongly influenced by magnetic geometry—can be engineered to enhance the inward pinch~[\onlinecite{10.1063/1.2789988}]. Another promising direction involves optimizing the denominator of the thermodiffusive term, which depends on the mode frequency, the drift frequency, and, when including parallel dynamics, also on parallel streaming, which is proportional to the parallel wavenumber \( k_\parallel \)~[\onlinecite{Angioni_2009}]. In particular, increasing \( k_\parallel \) to resonate with the frequency of the slow-passing electron population may further enhance inward transport. This can be achieved by increasing the local magnetic shear or by increasing the connection length through varying the number of field periods~[\onlinecite{10.1063/1.4868412}, \onlinecite{Plunk_PRL_2019}]. Together, these strategies define a broader class of stellarator optimization approaches, grounded in the physics of particle transport.

\textit{Acknowledgements.} The authors would like to thank A.~Zocco, F.~Wilms, G.~Merlo, D.~Fernando, and R.~Ramasamy for useful discussions. Numerical simulations were performed on the Marconi Fusion supercomputer at CINECA (Italy), the Viper cluster at the Max Planck Computing and Data Facility (MPCDF, Germany), and the LUMI supercomputer at CSC (Finland)


%

\end{document}